\documentclass[fleqn,usenatbib]{mnras}

\usepackage{newtxtext,newtxmath}

\usepackage[T1]{fontenc}

\DeclareRobustCommand{\VAN}[3]{#2}
\let\VANthebibliography\thebibliography
\def\thebibliography{\DeclareRobustCommand{\VAN}[3]{##3}\VANthebibliography}
\newcommand{\ltsimeq}{\raisebox{-0.6ex}{$\,\stackrel{\raisebox{-.2ex}{$\textstyle <$}}{\sim}\,$}}
\newcommand{\gtsimeq}{\raisebox{-0.6ex}{$\,\stackrel{\raisebox{-.2ex}{$\textstyle >$}}{\sim}\,$}}
\newcommand{\rvir}{R_{\rm vir}}
\newcommand{\fV}{f_{{\rm V}}}

\newcommand{\Rimp}{S_{\perp}}

\usepackage{graphicx}	
\usepackage{amsmath}	

\title[Molecular gas in the CGM]{ALMACAL X: Constraints on molecular gas in the low-redshift circumgalactic medium}

\author[A. Klitsch et al.]{Anne Klitsch,$^{1}$\thanks{E-mail: anne.klitsch@gmail.com}
Timothy A. Davis,$^{2}$\thanks{E-mail: DavisT@cardiff.ac.uk}
Aleksandra Hamanowicz,$^{3}$
Freeke van de Voort,$^{2}$
C\'eline P\'eroux,$^{4}$
\newauthor
and Martin A.~Zwaan${^4}$
\\
$^{1}$DARK, Niels Bohr Institute, University of Copenhagen, Jagtvej 128, 2200 Copenhagen N, Denmark\\
$^{2}$Cardiff Hub for Astrophysics Research \&\ Technology, School of Physics \&\ Astronomy, Cardiff University, Queens Buildings, Cardiff, CF24 3AA, UK\\
$^{3}$ Space Telescope Science Institute, 3700 San Martin Drive, Baltimore, MD 21218, USA\\
$^{4}$ European Southern Observatory (ESO), Karl-Schwarzschild-Strasse~2, D-85748 Garching, Germany
}
\date{Accepted XXX. Received YYY; in original form ZZZ}

\pubyear{2022}

\begin{document}
\label{firstpage}
\pagerange{\pageref{firstpage}--\pageref{lastpage}}
\maketitle

\begin{abstract}
Despite its crucial role in galaxy evolution, the complex circumgalactic medium (CGM) remains underexplored. 
Although it is known to be multi-phase, the importance of the molecular gas phase to the total CGM mass budget is, to date, unconstrained. 
We present the first constraints on the molecular gas covering fraction in the CGM of low-redshift galaxies, using measurements of CO column densities along sightlines towards mm-bright background quasars with intervening galaxies.
We do not detect molecular absorption against the background quasars. 
For the individual, low-redshift, `normal' galaxy haloes probed here, we can therefore rule out the presence of an extremely molecular gas-rich CGM, as recently reported in high-redshift protoclusters and around luminous active galactic nuclei. We also set statistical limits on the volume filling factor of molecular material in the CGM as a whole, and as a function of radius. ISM-like molecular clouds of $\sim$30\,pc in radius with column densities of N(CO)~$\gtsimeq 10^{16}$ cm$^{-2}$ have volume filling factors of less than 0.2 per cent. Large-scale smooth gas reservoirs are ruled out much more stringently. 
The development of this technique in the future will allow deeper constraining limits to be set on the importance (or unimportance) of molecular gas in the CGM.
\end{abstract}





\section{Introduction}

The cosmic cycle of multiphase baryons is one of the critical processes shaping the evolution of galaxies. Inflows of gas refuel the galaxy's reservoir for star formation, which, in turn, drives strong outflows of multiphase gas out of the galactic disc through stellar winds and supernovae \citep{Tumlinson2011, Tripp2011}. 
Most of the gas in this cycle resides in the medium surrounding the galaxy called the circumgalactic medium (CGM). 
The extensive, loosely defined, multiphase gaseous haloes host a substantial fraction of the baryons in the Universe \citep[e.g.][]{Shull2012}. 

Despite its crucial role for galaxy evolution, the CGM is difficult to fully characterise, due to its low surface brightness \citep[e.g.][]{Augustin2019} and its wide range of temperatures and densities requiring multiwavelength observations. 
The CGM is studied primarily through absorption lines in the spectra of background quasars. 
Yet current observational studies of these absorbers cover primarily the warm and hot phase of the CGM, represented by the hydrogen Ly$\alpha$ and UV metal absorption lines \citep[e.g.][]{Tumlinson2017, Peroux2020}. 
The cold gas phase of the CGM remains poorly constrained. 

Simulations show that the CGM is highly complex. 
Inflows from the intergalactic medium and outflows driven by, e.g., supernovae and active galactic nuclei interact, potentially guided by magnetic fields and cosmic rays \citep[e.g.][]{Peeples2019, Ji2020, vandeVoort2021, Stern2021}.
These large-scale interacting flows are responsible for driving strong turbulence, distributing metals throughout the CGM, creating large density contrasts, and allowing cooler gas pockets to form. 
While it has long been clear that these dense gas pockets form atomic gas (as probed by ubiquitous Ly$\alpha$ absorption lines), it is not known if molecules can also form and survive in the CGM. 

While detecting CGM gas is always challenging, the molecular gas phase is especially elusive. 
It is unclear how best to search for molecular gas not associated with the extended discs of galaxies or recently stripped from satellites. A number of studies tried to detect molecular gas, via either H$_2$ or CO, in high-velocity or intermediate-velocity clouds around the Milky Way with varying levels of success \citep[e.g.][]{Wakker1997, Akeson1999, Richter1999, Richter2003, Tchernyshyov2022}. Such clouds are often found reasonably close to the Milky Way \citep[e.g.][]{Wakker2008} with high atomic hydrogen column densities. 
In the haloes of external galaxies, \citet{Muzahid2015} targeted Damped Lyman $\alpha$ absorbers (DLA) harboring warm H$_2$ gas observed through UV absorption lines. 
They studied a large sample of absorbers at various impact parameters, i.e. the projected distance between an absorber and the host galaxy. 
Based on the detection rate, which was uncorrelated with the impact parameter, they concluded that H$_2$-bearing absorbers do not trace the molecular gas disk of the intervening galaxies but pockets of dense molecular gas within the CGM. 
Similarly, \citet{Klitsch2021} studied the cold molecular gas content of DLA host galaxies that harbor warm H$_2$ gas via CO emission line observations. 
They found that H$_2$ molecular gas absorbers are associated with molecular gas-rich galaxy overdensities and that the absorbers trace pockets of molecular gas in the CGM.
However, these studies specifically targeted known atomic gas absorbers with very high column densities, which could be gas stripped from satellites and not the general environments of the host galaxies. 

At higher redshift, several protoclusters have been found that host very extended molecular gas reservoirs (with densities high enough for CO to survive) in their intra(proto)cluster medium. 
This gas is not associated with any known galaxy \citep[e.g.][]{Emonts2016, Ginolfi2017, Emonts2019, Cicone2021, DeBreuck2022}. Luminous active galactic nuclei (AGN) at high-$z$ have also been found to host extended molecular gas \citep[e.g.][]{Jones2023,2022arXiv221203270S}.
However, again the origin of such gas is unclear, as is if such gas is common in the CGM of individual galaxies in less extreme environments, and whether these extended molecular regions exist at low redshift. 

To fully understand the distribution of cold molecular gas and the multiphase nature of the CGM, we here target the haloes around a set of galaxies seen in projection in the foreground of mm-bright quasars, probing a broad distribution of both galaxy and absorber properties, including their impact parameters. 
We present the first untargeted study of the molecular gas content of the low-redshift CGM traced through $^{12}$CO absorption lines using a sample of 142 galaxy-quasar pairs. 
We note that by using CO absorption lines to study molecular gas conclusions drawn from this study only concern metal-enriched molecular gas.
In Section~\ref{sec:data} we describe our quasar sample. We present our results in Section~\ref{sec:results} and give upper limits for the CO column density in the CGM. We conclude in Section~\ref{sec:conclusions}.

\section{Data \& sample} \label{sec:data}

In this project we make optimal use of available archival data from the ALMACAL survey\footnote{\url{https://almacal.wordpress.com/}} \citep{Zwaan2022} cross-matched with galaxy catalogues as described below.

\subsection{Target selection}

We use data from the ALMACAL survey that utilises ALMA calibration observations taken as part of other science projects from Cycles 1 to 6, taken before 2018 December 4.
These targets are extra-galactic, millimetre bright quasars that are used as amplitude, bandpass and phase calibrators \citep{Bonato2018}. 
The frequency coverage of the calibration observations is set by the science observations and integration times are usually short ($\sim 5$min).
We use all ALMA calibrators from \citet{Klitsch2019} with known redshifts from an updated redshift catalogue (S. Weng private communication) and cross-matched these sightlines with galaxies with known redshifts from the Simbad\footnote{\url{http://simbad.u-strasbg.fr/simbad/}} \citep{Wenger2000} and NED\footnote{\url{https://ned.ipac.caltech.edu/}} databases. 
These ALMA calibrators are at a redshift of $0.02 < z < 2.4$ with the majority lying between $0.55 < z < 1.3$.
We find 492 galaxies that lie within a projected distance of 150\,kpc from the quasars and spanning the same redshift range. 
However, because of the incomplete frequency coverage of the spectra we can only measure molecular gas column densities for 142 galaxy-quasar pairs, which have a median z$\approx$0.05.
The vast majority of the haloes we probe here are thus in the low-$z$ universe, backlight by QSOs at significantly higher-$z$ (median $\Delta z$=0.85). We would expect additional galaxies to exist in the intervening space, however the  spectroscopic (and, to a lesser extent, photometeric) catalogues of these regions are incomplete and biased. Thus the true number of haloes probed by this study is expected to be higher. Dedicated searches for the redshifts of photometric sources in these QSO fields would greatly assist in this regard.

\subsection{Galaxy properties} \label{sec:stellarmass}

It is important to compare the projected impact parameters with the galaxies' virial radii to know whether we are likely probing the CGM of these galaxies. Only a small fraction of these galaxies have stellar masses available from the literature. 
To ensure a uniform treatment we choose to apply the same technique to all galaxies.
Therefore, we calculate galaxy stellar masses based on their Ks or W1 magnitudes, using Eq.~2 from \citet{cappellari2013}.

\begin{figure}
    \centering
    \includegraphics[width = \linewidth]{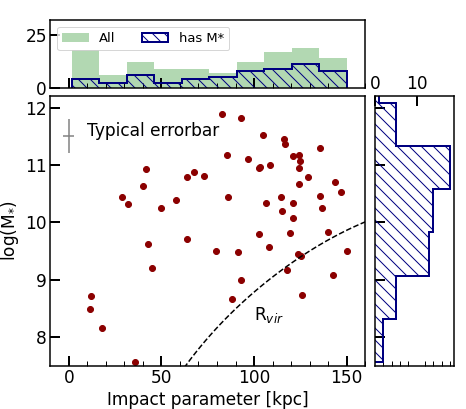}\vspace{-0.5cm}
    \caption{ Impact parameter and stellar mass of the galaxies near ALMACAL sightlines selected in this study. Stellar masses were estimated from the Ks/ W1 magnitudes (for 50 per cent of the sample). Histograms show the distribution of the corresponding parameter. The top histogram shows a distribution of impact parameters for all galaxies (green filled) and ones with derived stellar masses (blue hatched). A dashed black line marks the approximate virial radius for the corresponding stellar mass of the galaxy. In the top left corner, we show a typical errorbar of the datapoints. We find that the majority of sightlines trace gas within the virial radius.}
    \label{fig:m_star-impact-param_sample}
\end{figure}

The distribution of impact parameters and stellar masses are presented in Fig.~\ref{fig:m_star-impact-param_sample}. 
We estimate the virial radius using the stellar mass -- halo mass relation for redshift $z = 0$  from \citet{Legrand2019}.
We probe a range of impact parameters (10-140 kpc) and stellar masses (log(M*) $\sim$ 7-12), and the vast majority of the sightlines lie well within the virial radius of the host galaxy. 

\subsection{ALMACAL data reduction}

Here we only give a brief summary of the observations and data reduction.
For a detailed description of we refer the reader to \citet{Klitsch2019}. 

\citet{Klitsch2019} developed an optimised data reduction process for analysing a large number of spectra from the ALMACAL survey.
In short, we sum the XX and YY polarisation data that are extracted in the uv plane from the raw calibrator data. 
A bandpass correction is applied by taking the ratio of two calibrators observed in the same execution block.
Because of imperfections in this bandpass correction, additional masking of edge channels, atmospheric lines, and other bad data is applied. 
Finally, the remaining low frequency signal is subtracted from the spectra.

In their untargeted survey for intervening molecular absorption \citet{Klitsch2019} report no intervening absorption towards ALMA calibrators. 
We have analysed these data here as a pilot study, showing that this is a promising method to probe dense material in the CGM of galaxies.

\section{Analysis \& Results} \label{sec:results}

\subsection{Molecular gas column density}

We do not detect any CO absorption lines in the CGM of our target galaxies. 
Therefore, we derive $5\sigma$ upper limits on the CO column density for each intervening galaxy halo using the CO transition that would be covered by the spectrum given the redshift of the host galaxy. We apply the required corrections to extract the total column density using the excitation temperature and atomic constants as described by \citet{Mangum2015}. 
We assume an absorption line FWHM of 40~km~s$^{-1}$ \citep{Wiklind2018} and an excitation temperature equal to the cosmic microwave background temperature at the redshift of the host galaxy. 
The molecular column density depends linearly on the absorption line FWHM.
For a detailed discussion see \citet{Klitsch2019}.

In addition to these spectra from \citet{Klitsch2019} that we use in later sections, we have also stacked spectra covering the same frequency range to reach higher sensitivity, and check if any molecular signatures are likely to be hiding just below our detection thresholds. 
For this we have selected all spectra that cover a velocity range of $\pm 700 {\rm km \; s^{-1}}$. 
We choose this range based on the observed velocity difference between H$_2$ molecular gas absorbers and CO molecular gas-rich galaxies reported by \citet{Klitsch2021}.
Further we select only those spectra that have a velocity resolution of $< 10\,{\rm km \; s^{-1}}$ and smooth those to a common velocity resolution of $40\,{\rm km \; s^{-1}}$.
This yields a total number of 45 galaxies for which stacking is possible. 
We do not detect CO absorption in the stacked spectra. For each galaxy we calculate the CO column density from all available spectra (different CO transitions and stacks) and use the tightest constraint in the following analysis.

As an additional detection experiment we also stack all spectra covering the CN-A and CN-B lines at 113.494 85 GHz and 113.168 83 GHz. 
Because of the hyperfine structure of CN, these lines are expected to produce a broad absorption signal and are therefore a promising candidate for stacking experiments.
This stacking can be performed for 57 galaxies. 
However, we do not detect CN absorption in our stacked spectra.

Our limits span a range of column densities, as they are from fairly shallow calibration observations. However, we stress that the formation of CO molecules requires an environment shielded from UV light, either by dust or other molecules. CO thus should not exist at column densities $N(\mathrm{CO})\ltsimeq10^{13}$ cm$^{-2}$ \citep{2008ApJ...687.1075S}. As such our limits do probe much of the allowed parameter space. 

\begin{figure}
    \centering
    \includegraphics[width = \linewidth,trim=0.6cm 0.6cm 0.6cm 0.6cm,clip]{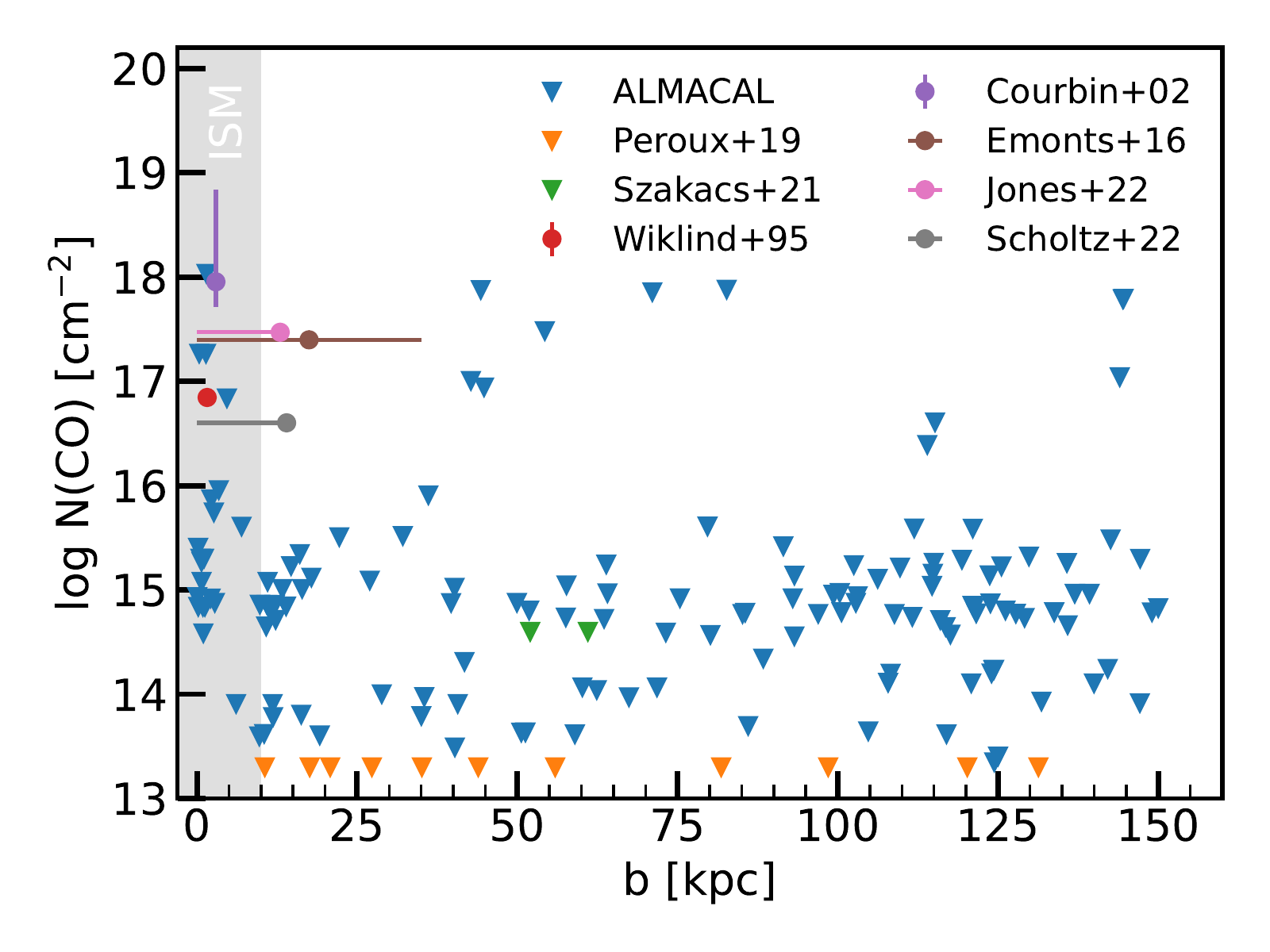}\vspace{-0.5cm}
    \caption{CO column density as a function of impact parameter. Sightlines at impact parameters $< 10$~kpc are marked in grey and excluded from further analysis as these may pass through the galaxy disk. In addition to our upper limits we show measurements and upper limits from the literature \citep{1995A&A...299..382W,2002ApJ...575...95C,Emonts2016, 2019MNRAS.485.1595P, Szakacs2021,Jones2023,2022arXiv221203270S}. For the haloes probed by our sightlines, we can rule out the presence of CGM molecular clouds with column densities similar to those detected in high-redshift protoclusters and high-redshift AGN.}
    \label{fig:nco-impact-param}
\end{figure}

In Fig.~\ref{fig:nco-impact-param} we show our column density limits against impact parameter as blue downward facing triangles. 
The region within a projected distance of $<10$\,kpc is marked in grey, as these sightlines may pass through ISM material in addition to the CGM. 
We also show the CO  detections of \citet{1995A&A...299..382W} (red circle), \citet{2002ApJ...575...95C} (purple circle), \citet{Jones2023} (pink circle), \citet{2022arXiv221203270S} (grey circle) and non-detections of \citet{2019MNRAS.485.1595P} (orange triangles) and \citet{Szakacs2021} (green triangles). 
We show the detection of extended CO in emission from the Spiderweb protocluster \citep{Emonts2016} as a brown circle, where we have assumed the material is uniformly distributed in a disc. 
If, as expected, this material is clumpy on some level then its true column density is likely to be higher. 
As can clearly be seen in Fig.~\ref{fig:nco-impact-param}, even our shallow limits from ALMACAL can rule out the presence of molecular clouds with similar column densities to those in the Spiderweb (and the other detected systems) along our lines of sight.

In what follows we use this fact to set limits on the column density and volume filling factor that dense clouds containing CO can have in the typical haloes probed by ALMACAL to still be consistent with our data.

\subsection{Molecular gas volume filling factor limits}

In order to derive volume filling factor limits we follow the formalism described by \cite{Stern2016}. Briefly, we assume the CGM is filled with homogeneous spherical clouds which can be 
characterized using the cloud column density N(CO), the cloud size $r_{c}$, and the volume filling factor $\fV$. 
We assume a spherically-symmetric CGM in which $\fV$ varies as a power law in $R$, and is zero beyond $\rvir$:
\begin{equation}\label{eq: fV def single rho}
\fV(R) = \fV(\rvir) \left(\frac{R}{\rvir}\right)^{l} ~~ \{R < \rvir\}.  
\end{equation}
As such we are left with five free parameters $(\rvir$, N(CO), $r_{c},\fV(\rvir),l)$ that define this idealized cold CGM model, and can be used to derive aggregate CGM quantities.


The expected number of clouds encountered within an individual sightline passing through the halo (also known as the covering factor $f_c$) can be calculated in this formalism as
\begin{equation}\label{eq: NH Rimp avg single rho}
f_c = \frac{3}{4r_{c}} \int^{\Rimp}_{-\Rimp} \fV(\sqrt{b^2+s^2})\, ds, 
\end{equation}
where $d s$ is the line element, $b$ the impact parameter and $\Rimp$ is the chord length ($\Rimp=\sqrt{\rvir^2-b^2}$ ). 
We can thus evaluate the expected number of clouds observed by ALMACAL, given the observed impact factor distribution, for any given $\fV$, $r_c$ and $l$. 
From the resulting covering fraction we can determine the probability that we obtain no detections for that set of parameters given $n$ draws from a Binomial distribution. 

\begin{figure}
    \centering
    \includegraphics[width = \linewidth,trim=0.6cm 0.6cm 0.6cm 0.6cm,clip]{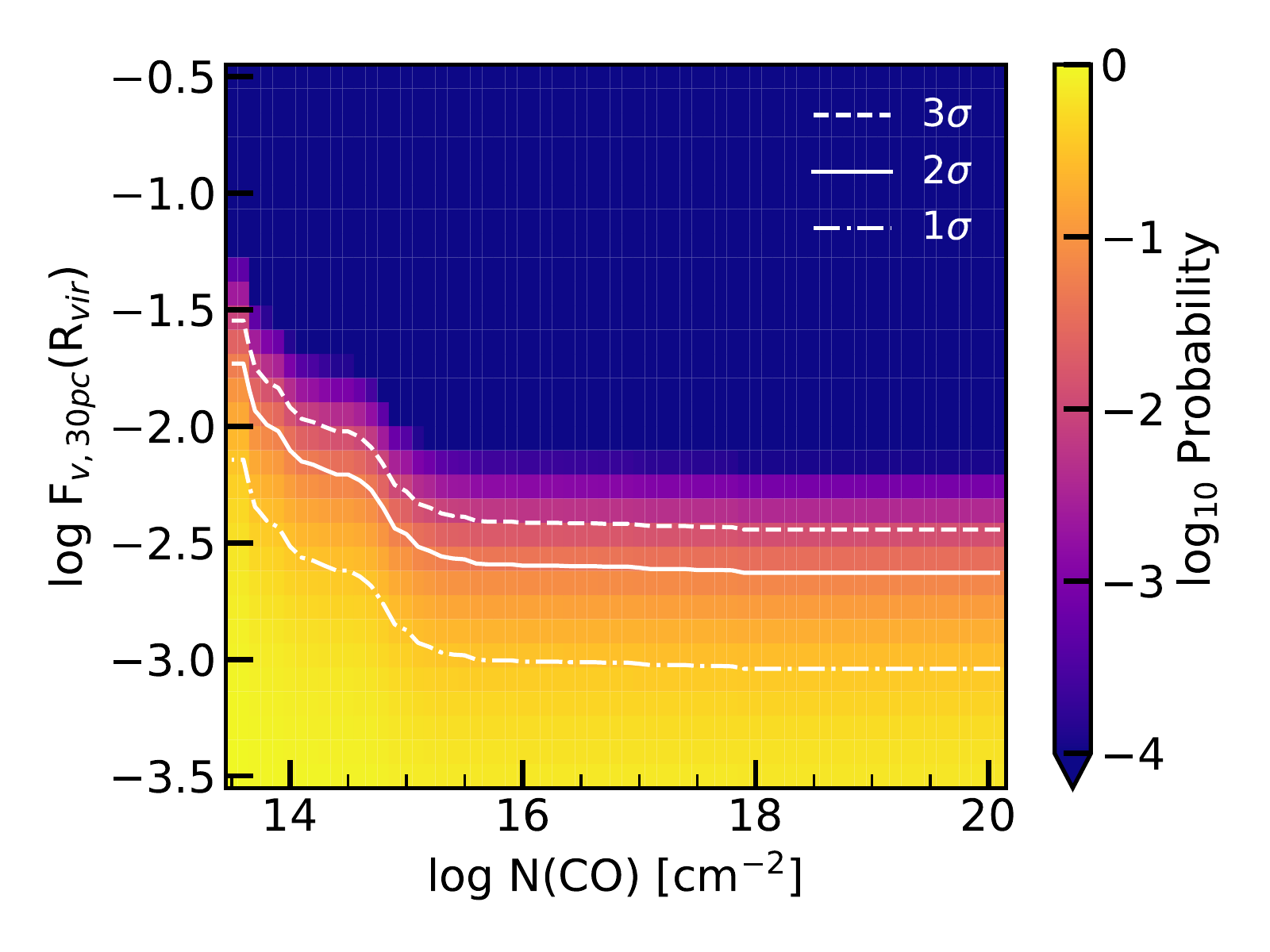}\vspace{-0.5cm}
    \caption{Global limits on the molecular gas volume filling fraction of CGM clouds with a given column density at fixed cloud size. The probability of observing no CGM clouds in the ALMACAL sample at a given column density and filling factor is shown as the colour-scale, with the 1, 2 and 3$\sigma$ limits highlighted with white dot-dashed, solid, and dashed lines respectively. We find that CGM molecular clouds at $\mathrm{N(CO)}>10^{16}$\,cm$^{-2}$ (for example) can only make up \ltsimeq0.075 per cent of the halo by volume. }
    \label{fig:volumeFillingFactor}
\end{figure}

In Fig.~\ref{fig:volumeFillingFactor} the colour-scale shows these probabilities as function of N(CO) and $\fV$. Here we have assumed an average $\rvir=150$\,kpc (approximately the median $\rvir$ expected for those objects in our sample with stellar mass measurements discussed in Section~\ref{sec:stellarmass}), that the mean $r_c=30$\,pc \citep{Farber22}, and that $l=-1$ \citep{Stern2016}. We discuss the impact of these assumptions below. We have excluded the sightlines with impact parameters smaller than 10\,kpc, as they may pass through the ISM, rather than purely through the CGM. White lines denote the 1, 2 and 3$\sigma$ limits we can place on the volume filling fraction of CO clouds in the CGM of our entire sample of ALMACAL galaxies. At 95 per cent confidence we find that $\fV(\rvir)$ for clouds with column densities $\mathrm{N(CO)}>10^{16}$\,cm$^{-2}$ is \ltsimeq0.25 per cent. Lower column density clouds ($\mathrm{N(CO)}>10^{14}$\,cm$^{-2}$) are less well constrained by the data ($\fV(\rvir)$\ltsimeq0.8 per cent).

\begin{figure}
    \centering
        \includegraphics[width = \linewidth,trim=0.6cm 0.6cm 0.6cm 0.6cm,clip]{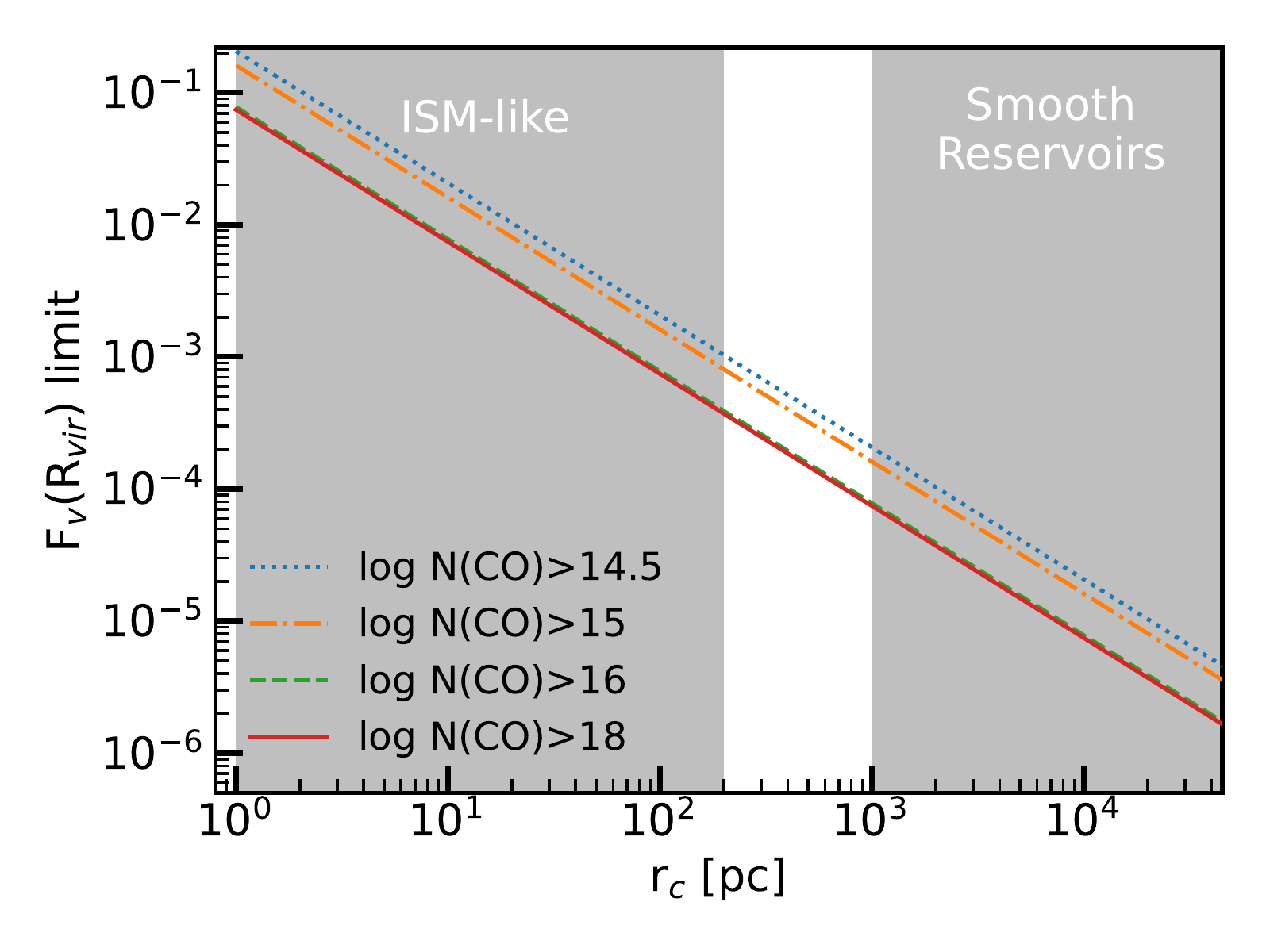}\vspace{-0.5cm}
    \caption{Global limits (at 95\% confidence) on the molecular gas covering fraction in the CGM in our sample galaxies, plotted as a function of assumed CGM cloud size. Smooth gas reservoirs, if they exist, can only fill a very small fraction of the halo of our typical low-z sources, and the most extreme smooth-reservoirs are essentially ruled out. }
    \label{fig:fillFactor_radius}
\end{figure}

The above values were calculated assuming a fixed molecular cloud size of 30\,pc. The derived column density limits scale linearly with the cloud size, as shown in Fig.~\ref{fig:fillFactor_radius}. For instance, if the CO hosting clouds in the CGM have a mean radius of 100\,pc, then at $\mathrm{N(CO)}>10^{16}$\,cm$^{-2}$ they can instead make up \ltsimeq0.075 per cent of the halo by volume. If instead the molecular gas is distributed over 10 kpc (as seen in some high-redshift sources) then it can make up less than \ltsimeq0.00075 per cent of the halo by volume on average. 

The existence of such large smooth reservoirs is considered unlikely, because gas that is dense enough to form CO should fragment \citep[and the gas conditions in the large reservoirs seen at high-$z$ seem to be more like those seen in the much smaller ISM clouds in our galaxy,][]{Emonts2019}. Encouragingly, such large smooth reservoirs are essentially ruled out by these data, as the volume enclosed by such a cloud would be much larger than 0.00075 per cent of the total halo volume. While this statement is somewhat complicated given that our analysis combines different sightlines through haloes of different masses, it seems likely that CO clouds in the CGM (if they exist) must be similar in size to typical ISM molecular clouds. 

\begin{figure}
    \centering
    \includegraphics[width = \linewidth,trim=0.6cm 0.6cm 0.6cm 0.6cm,clip]{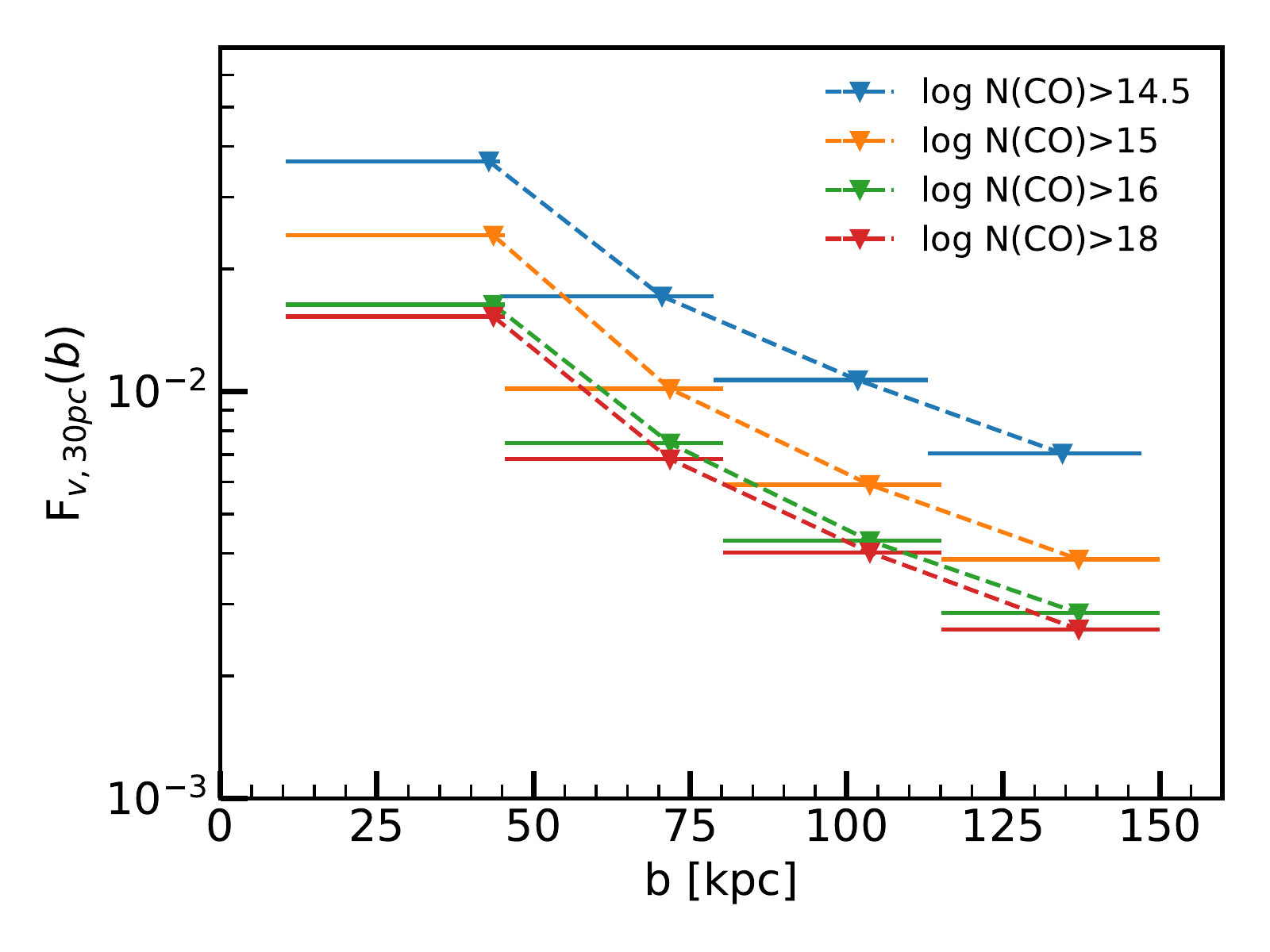}\vspace{-0.4cm}
    \caption{Limits on the molecular gas covering fraction in the CGM as a function of radius assuming cloud sizes of 30~pc. These limits are more or less constraining depending on the availability of measurements that probe the corresponding impact parameter bin (see Fig.~\ref{fig:nco-impact-param}). Clouds with column densities $\mathrm{N(CO)}>10^{16}$\,cm$^{-2}$ are constrained to take up \ltsimeq2 per cent of the halo within $\sim$45\,kpc, and \ltsimeq0.2 per cent within $\sim$130\,kpc.}
    \label{fig:volumeFillingFactor_radius}
\end{figure}

In Fig.~\ref{fig:volumeFillingFactor_radius} we split the halo into four radial bins, assuming ISM-like clouds 30~pc in size. Here we show the 95 per cent confidence interval constraint on the volume filling factor at the area weighted bin centroid (i.e $\fV(\bar{b})$ in each bin), rather than $\fV(\rvir)$, to better visualise the constraints we can set in different parts of the CGM. Because of the lower number of sightlines constraining the central bin, these constraints are weakest, while all our sightlines can be used to constrain the presence of CO gas clouds in the outer regions of the CGM.  Clouds with column densities $\mathrm{N(CO)}>10^{16}$\,cm$^{-2}$ are constrained to take up \ltsimeq2 per cent of the halo within $\sim$45\,kpc, and \ltsimeq0.2 per cent within $\sim$130\,kpc.

In all the above we have assumed a fixed $\rvir=150$\,kpc, and that $l=-1$, following \cite{Stern2016}. If the true virial radii of our sources are larger (smaller) then our filling factor constraints become slightly more (less) stringent (e.g. our constraints are $\approx$20\% stronger if $\rvir=200$\,kpc). Changing $l$ does not significantly impact our constraints on the inner halo, but does change the outer halo points in Fig.~\ref{fig:volumeFillingFactor_radius}; by a factor of $\approx$2 at $\approx$130\,kpc when allowing $l$ to change between $-1.5$ and $-0.5$. Although we do not know the true virial radius and value for $l$ these changes are relatively small, and would not change our conclusions.


\section{Discussion and conclusions} \label{sec:conclusions}

The gaseous haloes around galaxies are known to contain neutral gas as well as ionized gas, but there are few constraints on their molecular gas content. We used data from the ALMACAL survey to probe CO in the CGM of low-redshift galaxies. Our galaxy sample is diverse and spans a large range in stellar mass (from $10^{7.5}$ to $10^{12}$~M$_{\odot}$). Our quasar sample probes random sightlines around these galaxies, with impact parameters from 0 to 150~kpc, and was not selected based on any known absorption lines. The non-detection of CO in any of our quasar spectra allows us to put constraints on the low-redshift molecular CGM. We note that by using CO absorption lines to study molecular gas conclusions drawn from this study only concern metal-enriched molecular gas.

We place limits on the covering fraction of molecular gas that depend on the minimum column density and impact parameter. For our low-redshift sample of galaxies, we can rule out smooth molecular gas reservoirs with column densities similar to those observed in extreme protocluster environments and around luminous AGN at high redshift \citep[e.g.][]{Emonts2016,Jones2023}. In the inner CGM (between 10 and 45~kpc), our data rule out covering fractions of more than 2 per cent for gas with CO column densities above $10^{16}$~cm$^{-2}$. Out to $\sim130$~kpc, the contraints are even tighter and only covering fractions below 0.2 per cent are allowed by our data. However, note that, because this gas is very dense, a sizeable amount of molecular gas could still be hiding even at these low covering fractions.

Missing frequency coverage and short integration times are major limitations of this pilot study, both of which we could address with dedicated observations in the future.  Given the statistical nature of these constraints, the largest gains can be realised by substantially expanding the sample with observations at moderate depth, rather than small numbers of very deep observations of individual sightlines. Given that a limited number of millimetre-bright quasars exist, efforts to better characterise the galaxy population along existing lines-of-sight would also be valuable. 

\vspace{-0.5cm}
\section*{Acknowledgements}

We wish to thank the ALMACAL Team and in particular Simon Weng, Jianhang Chen, Elaine Sadler and Victoria Bollo for constructing the new redshift catalogue and useful discussions.
We would also like to thank Bjorn Emonts for very helpful discussions. 
A.K.~gratefully acknowledges support from the Independent Research Fund Denmark via grant number DFF 8021-00130. TAD acknowledges support from the UK Science and Technology Facilities Council through grants ST/S00033X/1 and ST/W000830/1.
FvdV is supported by a Royal Society University Research Fellowship (URF$\backslash$ R1$\backslash$ 191703).
ALMA is a partnership of ESO (representing its member states), NSF (USA), and NINS (Japan), together with NRC (Canada), NSC and ASIAA (Taiwan), and KASI (Republic of Korea), in cooperation with the Republic of Chile. The Joint ALMA Observatory is operated by ESO, AUI/NRAO, and NAOJ.
This work made use of Astropy\footnote{\url{http://www.astropy.org}} a community-developed core Python package for Astronomy \citep{astropy2022}
and Astroquery \citep{Ginsburg2019}.
This research has made use of the SIMBAD data base, operated at CDS, Strasbourg, France. 
This research has made use of the NASA/IPAC Extragalactic Data base (NED), which is operated by the Jet Propulsion Laboratory, California Institute of Technology, under contract with the National Aeronautics and Space Administration.
This research has made use of NASA's Astrophysics Data System.

\vspace{-0.5cm}
\section*{Data Availability}

Raw data used in this work is available from the ALMA science archive (\url{https://almascience.eso.org/}). 
Processed data will be shared upon reasonable request.
\vspace{-0.5cm}











\bsp	
\label{lastpage}
\end{document}